\documentclass[11pt,letterpaper]{article}
\usepackage[utf8]{inputenc}
\usepackage[top=1in, bottom=0.75in, left=0.75in, right=.75in]{geometry} 
\usepackage{graphicx}   
\usepackage{enumitem}  
\usepackage{amsmath}
\usepackage{cite}
\usepackage[numbers,sort&compress]{natbib}
\usepackage{natbibspacing}

\newcounter{affno}
\renewcommand\theaffno{\alph{affno}}
\newcommand{\aff}[1]{\refstepcounter{affno}\label{#1} \textsuperscript{\theaffno}}
\newcommand{\aref}[1]{\textsuperscript{\ref{#1}}}
\newcommand{\arefs}[2]{\textsuperscript{\ref{#1},\ref{#2}}} 

\newcommand\snowmass{\begin{center}\rule[-0.2in]{\hsize}{0.01in}\\\rule{\hsize}{0.01in}\\
\vskip 0.1in Submitted to the  Proceedings of the US Community Study\\ 
on the Future of Particle Physics (Snowmass 2021)\\ 
\rule{\hsize}{0.01in}\\\rule[+0.2in]{\hsize}{0.01in} \end{center}}

\begin{document}

\title{Lattice QCD and the Computational Frontier}

\author{Peter Boyle\arefs{BNL}{UoE},
Dennis Bollweg\aref{CU},
Richard Brower\aref{BU},
Norman Christ\aref{CU},
Carleton DeTar\aref{UTAH},\\
Robert Edwards\aref{JLAB},
Steven Gottlieb\aref{Indiana},
Taku Izubuchi\arefs{BNL}{RBRC},
Balint Joo\aref{ORNL},
Fabian Joswig\aref{UoE},\\
Chulwoo Jung\aref{BNL},
Christopher Kelly\aref{CSI},
Andreas Kronfeld\aref{FNAL},
Meifeng Lin\aref{CSI},\\
James Osborn\aref{ANL},
Antonin Portelli\aref{UoE},
James Richings\aref{UoE},
Azusa Yamaguchi\aref{UoE}
}

\date{\today}

\bibliographystyle{JHEP}

\maketitle

\subsection*{\textsf{Affiliations}}
{\small \flushleft
\aff{BNL} Physics Department, Brookhaven National Laboratory, Upton, NY 11973, USA\\
\aff{UoE} Higgs Centre for Theoretical Physics, The University of Edinburgh, EH9 3FD, UK\\
\aff{CU} Physics Department, Columbia University, NY 10024, USA\\
\aff{BU} Department of Physics, Boston University,590 Commonwealth Avenue, Boston, MA 02215, USA\\
\aff{UTAH} Department of Physics and Astronomy, University of Utah, Salt Lake City, Utah 84112, USA\\
\aff{JLAB} Thomas Jefferson National Accelerator Facility, Virginia, USA\\
\aff{Indiana} Department of Physics, Indiana University, Bloomington, Indiana 47405, USA\\
\aff{RBRC} RIKEN BNL Research Center, Brookhaven National Laboratory, Upton, NY 11973, USA\\
\aff{ORNL} Oak Ridge National Laboratory, Oak Ridge, TN 37831, USA\\
\aff{FNAL} Theory Division, Fermi National Accelerator Laboratory, Batavia, IL 60510, USA\\
\aff{CSI} Computational Science Initiative, Brookhaven National Laboratory, Upton, NY 11973\\
\aff{ANL} Argonne National Laboratory, Lemont, IL 60439\\
}

\subsection*{\textsf{Abstract}}

The search for new physics requires a joint experimental and theoretical effort.
Lattice QCD is already an essential tool for obtaining precise model-free theoretical
predictions of the hadronic processes underlying many key experimental searches,
such as those involving heavy flavor physics, the anomalous magnetic moment of the muon,
nucleon-neutrino scattering, and rare, second-order electroweak processes.
As experimental measurements become more precise over the next decade,
lattice QCD will play an increasing role in providing the needed matching
theoretical precision. Achieving the needed precision requires simulations with
lattices with substantially increased resolution.  As we push to finer lattice spacing we encounter an
array of new challenges. They include algorithmic and software-engineering challenges, challenges in
computer technology and design, and challenges in maintaining the necessary human resources.
In this white paper we describe those challenges and discuss ways they are being dealt with.
Overcoming them is key to supporting the community effort required to deliver the needed
theoretical support for experiments in the coming decade.


\snowmass

\pagebreak

\section{Introduction}

The USQCD lattice collaboration wrote a computing white paper in 2019~\cite{Joo:2019byq}. This Snowmass white paper is intended to complement and add to 
the 2019 document by anticipating longer-term technology trends
and the algorithmic and software development required.

The precise theoretical prediction of hadronic properties is a central part of searches for new physics over the coming decades.
Lattice gauge theory is a systematically improvable theoretical
tool for numerical evaluation of the Euclidean Feynman-path integral for
Quantum Chromodynamics (QCD) in its nonlinear regime
where it describes the hadrons as bound states of quarks.
Various discretization approaches are taken, which respect or break continuum flavour and chiral symmetries to differing degrees.
The lattice-gauge-theory simulation effort 
takes place worldwide and directly supports high-energy 
physics experiment by providing calculations of the properties
of hadrons that are in many cases vital to the interpretation of experimental results
and to the comparison of our observations with standard model predictions, thus
enabling searches for new physics. A comprehensive 
review of important lattice predictions for a range
of important quantities is carried out regularly by the
Flavour Lattice Averaging Group~\cite{Aoki:2021kgd}.

There are significant challenges to meeting the precision goals of the next decade and
beyond. These include the development of new algorithms and software that enable substantially greater numbers of degrees of freedom to be simulated on emerging 
supercomputers over the next 5 to 10 years.
This work is required in order to reduce theoretical uncertainties by enabling calculations with larger physical lattice volumes,
inclusion of both QCD and QED effects, finer lattices, and larger statistical samples.  Finer lattices also give access to substantially higher energy scales. 

Lattice QCD is already playing an important role in determining the
hadronic contributions to the anomalous magnetic moment of the muon, discussed
in a Snowmass white paper~\cite{MuonSnowmass}, by the international Muon $g-2$ Theory Initiative white paper~\cite{Aoyama:2020ynm} and in a white paper by the USQCD collaboration in 2019~\cite{USQCD:2019hyg}.
These calculations will be critical to the interpretation of results from the Fermilab Muon $g-2$ Experiment~\cite{Muong-2:2021ojo}, which is currently in excellent agreement with results from the earlier Brookhaven experiment~\cite{Bennett:2006fi}.

Understanding the neutrino-nucleus interaction is critical to the 
analysis of the Deep Underground Neutrino Experiment.
The interaction between an intermediate W or Z boson
with a quark confined inside a neutron or proton within the nucleus is theoretically
complex. This was discussed by USQCD in a 2019 white paper~\cite{Kronfeld:2018qcd}.

A robust quark-flavor physics program
has also been discussed in a USQCD white paper~\cite{USQCD:2019hyg}. 
These phenomena are typically highly suppressed in the standard model and therefore also offer promising avenues for the discovery of new physics.
Contributed Snowmass white papers include improving the
understanding of anomalies in B physics~\cite{OliverSnowmass}, 
emerging from the the Large Hadron Collider and studied further
at Belle II, as well as reconciling CP violation observed in kaon experiments with the standard model and exploring rare Kaon decays~\cite{NormanSnowmass}.

Simulation at finer lattice spacing and larger volumes introduces new challenges. In this white paper we analyze the array of new challenges facing lattice QCD in the next ten years. Meeting them requires new algorithmic research, novel computer hardware design beyond the exascale, improved software engineering, and attention to maintaining human resources.  If these challenges are met, lattice QCD will deliver the theoretical precision needed for interpreting experimental measurements in the foreseeable future.  We begin with algorithmic challenges in Sec.~\ref{Sec:algorithms} and provide a detailed discussion of the arithmetic intensity and heirarchical bandwidth requirements intrinsic to the various fermion formulations in use. We also provide an illustrative estimate of computational costs for problems that extend beyond the exascale.  We follow in Sec.~\ref{Sec:computational-landscape} with a discussion of the challenges in designing computer systems that address the computational needs. We go into some detail reviewing new memory technology and interconnect design. That review suggests further directions for algorithmic research. We discuss software-engineering challenges in Sec.~\ref{Sec:software-engineering}, offering suggestions for DOE influence in improved software tools. Finally, we discuss the human-resource challenges in Sec.~\ref{Sec:ecosystem-challenges}.

\section{Lattice QCD and Algorithms}
\label{Sec:algorithms}

In this section we introduce, briefly, the standard method for simulating QCD on a lattice and then turn to algorithmic challenges, computational requirements, and a discussion of estimated problem sizes and simulation costs in the coming years.

Lattice gauge theory formulates the Feynman path integral for QCD as a statistical
mechanical sampling of the related Euclidean space path integral. The sampling is performed
by Markov chain Monte Carlo (MCMC) sampling, using forms of the hybrid Monte Carlo (HMC) algorithm\cite{Duane:1987de,Clark:2006fx,Luscher:2004pav,Cossu:2017eys,Nguyen:2021zgx,Foreman:2021ljl}. 
The partition function of QCD is sampled through the introduction
of auxiliary momentum ($\pi$) and pseudofermion ($\phi$) integrals,
\begin{equation}
Z = 
  \int d\pi
  \int d \phi
  \int d U \quad
  e^{-\pi^2/2}
  e^{-S_G[U]}
  e^{-\phi^\ast (D^\dag D)^{-1} \phi }.
  \label{eq:Z}
\end{equation}
The sampling is weighted with the action probability functional, in which $S_G$ is the
gauge action and the sparse matrix $D$ represents the lattice Dirac operator.
The pseudofermion term provides a stochastic estimate of the fermion determinant 
which describes the effect of quark loops in the partition function.
It involves the inverse of the gauge-covariant Dirac operator and requires  linear-solver algorithms to evaluate.
The momentum integral serves only to move the gluon fields ($U$) efficiently around
their group manifold at constant energy in the HMC sampling algorithm.
As computers become more capable, we are able to simulate at smaller lattice spacings with higher precision, which increases the range of length scales covered.

Present algorithms for both MCMC sampling and Dirac solvers display growing limitations as substantially greater ranges of energy scales are included in our problem, an algorithmic challenge called critical slowing down.
Also for the computation of correlation functions the inversion of the matrix $D$ is typically the computationally most expensive component. While a greater degree of trivial parallelism can be utilized for these quark correlation
function calculations, a large number of inversions per gauge configuration is needed to achieve sub-percent precision for complicated observables.

When QCD is naively discretized, additional unphysical degrees of freedom appear associated with zero modes arising at the corners of the Brillouin zone in momentum space. Two techniques are commonly employed to control this problem: in the staggered fermion approach~\cite{HISQ} the path integral weighting is adjusted to account for the additional degrees of freedom ("tastes") and their contributions to the valence sector are handled theoretically, for example using staggered chiral perturbation theory. A second approach introduces a ``Wilson term'' which adds a large mass to the extra degrees of freedom proportional to the inverse lattice spacing, decoupling them from the theory in the continuum limit. Unfortunately the Wilson term breaks the chiral symmetry, which is the symmetry of QCD under the interchange of left and right-handed quark fields, and which is vital to the correct description of the pions that encapsulate the majority of the low-energy degrees of freedom in the system. The domain-wall-fermion approach~\cite{Shamir:1993zy} reintroduces the chiral symmetry in a controlled way by placing the chiral modes on opposite sides of a fictional, finite fifth dimension, at the expense of additional computational cost. Both the staggered and domain wall discretizations are employed extensively by USQCD in High Energy Physics (HEP) calculations.

\subsection{Algorithmic Challenges}
\label{sec:latmeth}

The development of numerical algorithms is an intellectual activity that  spans physics, mathematics, and computer science. Many key algorithms such as Markov Chain Monte Carlo (MCMC) and Metropolis-Hastings algorithms emerged from theoretical physics. The hybrid/Hamiltonian Monte Carlo algorithm was developed in Lattice Gauge theory and is one of several work-horse algorithms used in training machine learning.
Over the almost 40 year history of active lattice gauge theory calculations, the annual improvement purely 
from algorithm development has been broadly similar to, and 
additional to, the gain from Moore's law. 

The needed simulations cover a much greater range of length scales and energy scales than those to date. 
This will require a new generation of algorithms
tailored to multi-scale dynamics. In the absence of these algorithms
both Dirac-matrix inversion and Monte-Carlo sampling are known
to require increasingly large iteration counts or Monte-Carlo evolution
times as the lattice spacing is reduced. 
We hope that these issues may be addressed in one of several ways.

Firstly, the Atiyah-Singer index theorem guarantees the cost of Dirac matrix solution
becomes large as the continuum limit is taken. Krylov solvers are
fundamentally polynomial approximations to the inverse $f(x)=1/x$.
The order of the polynomial required for a given accuracy grows linearly
in the ratio of the highest to lowest eigenvalue (condition number) of the squared Dirac operator. 
This dictates a loss of algorithmic efficiency as the
lattice is made finer and the Nyquist frequency is increased. Therefore the Krylov solver takes a 
diverging number of iterations as the continuum limit is taken.

This difficulty can be countered in several ways. Firstly one
can accurately compute a number of the lowest lying eigenvectors of the 
Dirac operator and handle these 'exactly' up to numerical tolerance.
This removes these modes from a \emph{deflated} Krylov solver and changes the effective condition number, thus accelerating convergence.

Alternative multigrid~\cite{Brannick:2007ue,Luscher:2007se,Babich:2010qb} solvers
have demonstrated order-of-magnitude gains for Wilson fermions by
approximately and repeatedly handling degrees of freedom in the low
lying eigenspace as a form of preconditioner.
The US HEP program is presently focused on the
domain-wall\cite{Shamir:1993zy} and staggered approaches,
but corresponding gains for staggered~\cite{Brower:2018ymy} and domain-wall-fermion~\cite{Cohen:2011ivh,Boyle:2014rwa,Yamaguchi:2016kop,Boyle:2021wcf,Brower:2020xmc} discretizations
are an open research activity. 
In all cases gauge invariance dictates that the coarse degrees
of freedom must be discovered in a data-dependent way to 
treat more efficiently longer distance or collective degrees of freedom. This
discovery or setup step is costly and limits the gains from multigrid
in HMC. Better methods are required, and it is possible both eigenvector solution and multigrid setup
could be phrased as a well posed potential machine-learning feature-recognition problem.

A second direction is the critical slowing down of
MCMC algorithms. This manifests itself, among 
other metrics, as the freezing of topological charge in simulations
with periodic gauge field boundary conditions. 
An underlying issue is that the HMC forces for 
different ``wavelength'' modes in the system receive vastly different
forces in our fictitious Monte-Carlo time evolution. 
Several research activities are being undertaken to introduce 
a wavelength-dependent evolution speed in the Monte Carlo that
accelerates the sampling of long-distance dynamics to evade
critical slowing down. The first of these is the 
Riemannian Manifold Hybrid Monte Carlo (RMHMC) algorithm~\cite{Girolami2011}, which is in some
aspects similar to a ``gauge
invariant Fourier Acceleration'' idea introduced in ref.~\cite{Duane:1986fy}. This is being studied under Exascale Computing and
SciDAC projects, with recent progress~\cite{Nguyen:2021zgx}
and also in a related gauge-fixed form~\cite{Sheta:2021hsd},
suggesting that critical slowing down may be substantially reduced and possibly eliminated on small physical volumes.

Other algorithmic directions with different ways of attacking the
same problem include applications of machine learning
to configuration sampling. This possibility has
been discussed in a dedicated Snowmass white paper~\cite{Boyda:2022nmh}.

\subsection{Lattice QCD as a computational problem}

In this subsection we analyze the key characteristics of the computational problem that present challenges to computer hardware design.  In particular, we focus on the arithmetic intensity of the various fermion discretizations.

From a computational perspective Lattice QCD is performed on structured 
Cartesian grids with a high degree of regularity and natural data 
parallelism. The central repeated operation is the solution of the
gauge covariant Dirac equation. 
In a discrete system, the partial  derivative is replaced by a finite difference
with spacing $a$
\begin{equation}
(\partial_\mu - i g A_\mu(x))\psi(x)\rightarrow  \frac{ U_\mu(x) \psi(x + a \hat \mu)  -    U^\dag_\mu(x -a ) \psi(x - a\hat \mu)}{2 a}.
\end{equation}
Wilson's covariant displacement preserves the gauge
symmetry of QCD and is achieved with  
SU(3) matrices $U_\mu(x) = e^{ i a g A_\mu(x)}$ connecting lattice sites called gauge links.
These $3\times 3$ complex valued matrices represent the QCD
generalization of the electromagnetic vector potential acting
on color degrees of freedom.

The Dirac differential operator acts on a space-time field with (typically) 
12 degrees of freedom at each coordinate site. It is represented as a sparse 
matrix on the structured grid
where it typically has either 8 (or 16 points) entering the list of non-zero
entries in each row, with a nearest-neighbour (or next-next-nearest neighbour) 
'cross' geometry in four dimensions and with coordinate-dependent coefficient matrices
multiplying each contribution. 
The solution is performed using iterative Krylov solvers,
either in a standard algorithm such as conjugate gradient or with a multigrid
preconditioner. 

The Lattice QCD workflow is divided into two phases.
First, a MCMC sampling phase generates
an ensemble of the most likely gluon field configurations distributed
according to the QCD action.
The fields $U_\mu$ are modified throughout the MCMC,
so that algorithmic accelerations like multigrid or eigenvector
deflation require fresh setup or discovery
of the most relevant degrees of freedom. 
The ensemble generation is serially dependent and represents a strong scaling
computational problem. Ideally one would be able to use efficiently O($10^4$) computing nodes
on O($256^4$) data points. On the largest scales this becomes a halo-exchange
communication problem since the local data bandwidths vastly exceed those of 
inter-node communication.

In the second phase hadronic observables are calculated
on each sampled configuration where many thousands of quark propagators
are calculated and assembled into correlation functions.
This both allows more scope for amortizing
the setup cost of advanced algorithms like multigrid or deflation, and also has a high
degree of trivial parallelism. Computer interconnect limitations are more easily avoided  by exploiting this trivial parallelism. 
The local-memory bandwidth requirements of calculating quark propagators
for hadronic correlation functions can be
amortized and the computational rate can be brought closer to the 
cache-bandwidth limit through use of multiple right hand
side solvers and some trivial-level parallelism
between computing nodes.

We will consider the computational requirements of Wilson, Improved Staggered and domain wall fermion (DWF) actions and with both
single right-hand-side and multiple right-hand-side Krylov solvers.
Multiple right-hand-side solvers can be advantageous, where applicable, for Wilson and Staggered
actions because the gauge field is a significant element of the memory traffic, but it can be applied to multiple
vectors concurrently, suppressing this overhead.
The same suppression factor already occurs quite naturally for five-dimensional chiral fermion actions such as DWF.
When this natural gain is combined with solvers for multiple right-hand sides, 
the gain is amplified and both algorithmic and execution
performance can be accelerated.
We assume an $L^4$ local volume and either an  8-point (nearest neighbour) or 16-point (nearest neighbor + Naik) stencil.
Under the assumption of weak scaling, the local floating-point, 
memory and cache bandwidths and MPI-network bandwidth per node suffice to describe large systems. 

For either multiple right-hand sides (multiRHS) 
or for DWF, we  take the fifth
dimension size $L_s \equiv N_{\rm rhs}$.
A cache-reuse factor $\times N_{\rm stencil}$ for fermion fields 
is possible, but only if the cache is of sufficient capacity.
We can count the words accessed per 4d site 
by each node, 
and the surface volume that must be communicated
in a massively parallel simulation with this local volume. 
Here $N_d$ is the number of dimensions (excluding the fifth dimension for DWF), and $N_s$ and $N_c$ are the number of spin and color degrees of freedom, respectively.
These comprise
\begin{itemize}
\itemsep-0.3em
\item Fermion: $N_{\rm stencil} \times (N_s \in \{ 1,4 \}) \times (N_c=3) \times (N_{\rm rhs} \in \{ 1, 16 \})$ complex,
\item Gauge: $2 N_d \times N_c^2$ complex.
\end{itemize}
Similarly we can count the floating point operations as proportional to the number of points in the stencil $N_{\rm stencil}$
and the number of fermion spin degrees of freedom per point after (any) spin projection $ N_{hs}$. The
cost of spin projection, and the cost of SU(3) matrix multiplication is
\begin{itemize}
\itemsep-0.3em
\item  SU(3) MatVec: $66\times N_{\rm hs} \times N_{\rm stencil}$ .
\end{itemize}
We tabulate the arithmetic intensity and surface to volume ratio of the different kernels in Table~\ref{tab:arithint}.
\begin{table}[hbt]
\begin{tabular}{c|cccccccc}
Action  & Fermion Vol & Surface & $N_s$ & $N_{hs}$ & $N_{\rm rhs}$ & Flops & Bytes & Bytes/Flops\\
\hline
Staggered    & $L^4$ & $3\times 8\times L^3$ &  1  & 1 & 1 & 1146 & 1560 & 1.36  \\
Wilson  & $L^4$ & $8\times L^3$ &  4  & 2&  1 & 1320  & 1440 & 1.09      \\
DWF     & $L^4\times N$ & $8\times L^3$ &  4 & 2 & $L_s$ & $L_s \times 1320$ & $N_{\rm rhs}\times 864$ & 0.65\\
\hline
Wilson-RHS  & $L^4$ & $8\times L^3$ &  4  & 2 & $N_{\rm rhs}$ & $N_{\rm rhs} \times 1320$ & $N_{\rm rhs}\times 864$ &0.65 \\
Staggered-RHS    & $L^4$ & $3\times 8\times L^3$ &  1  & 1 & $N_{\rm rhs}$& $N_{\rm rhs}\times 1146$ & $N_{\rm rhs}\times 408$& 0.36\\
\end{tabular}
\caption{\label{tab:arithint}
Arithmetic intensity is the ratio of the 
floating point instructions executed to the bytes accessed
for each of the cache, memory and interconnect subsystems.
The arithmetic intensities (single precision) and surface to volume ratio for different forms of the QCD action and solver.
Here there is significant possible cache reuse on the
data due to the stencil nature, however this is only
possible with an idealized large-capacity cache.
Note that for Wilson and
staggered fermions the arithmetic intensity can be greatly improved for valence analysis by increasing the number of right-hand sides as indicated by the last two lines of the table.
}
\end{table}

Since for multiRHS Wilson or for DWF $\sim \frac{1}{L}$ of the fermion data comes from off node, scaling the fine operator requires a bidirectional interconnect bandwidth,
\begin{equation}
    B_{\mathrm{network}} \sim 
 \frac{2 N_{hs} }{N_s}
\frac{B_{\mathrm{memory}}}{L} \times R_{\mathrm{eff}} 
\label{eq:cacheReff}
\end{equation}
where $R_{\mathrm{eff}}$ is the effective \emph{reuse} factor obtained for the stencil in caches (assuming that the memory bandwidth $B_{\mathrm{memory}}$ is saturated).
The product $B_{\mathrm{memory}} \times R_{\mathrm{eff}}$ represents the actual
data throughput from the entire memory system, including all
levels of cache, and this \emph{actual} cache access rate can be empirically
measured using the arithmetic intensity and a measured single node
performance in flop/s via,
$$B_\mathrm{cache} = R_\mathrm{eff} B_\mathrm{memory} = (\mathrm{flops/second}) \times (\mathrm{Bytes/flop}).$$
It is now common for performance counters and profiling to measure $B_\mathrm{memory}$ and sometimes $B_\mathrm{cache}$
in isolation, giving several ways to determine the cache reuse rate $R_\mathrm{eff}$ and check that memory
bandwidth is saturated.
The network bandwidth should be sufficient to perform all transfers in the same amount of time as the computation, and can be
overlapped with interior terms in the computation that represent the majority of the calculation.
Equation~\ref{eq:cacheReff} reflects a factor of two (send + receive bidirectional bandwidth),
and spinor projection compression (Wilson, DWF) and a geometrical $\frac{1}{L}$ for
a nearest neighbour halo depth of size one.
This argument and the above expressions are easy to generalise to multiRHS HISQ fermions, which
acquire a $3/L$ factor due to the larger stencil.

As a concrete example,  a four-GPU computer node delivering a measured 10 TFlop/s
performance on single node DWF action with a $32^4$
local volume obtains an L2 cache throughput of
$$
B_\mathrm{cache} = 10000 \times 0.65 = 6500 \mathrm{GB/s}.
$$
This implies a per-node aggregate interconnect throughput
requirement of 200 GB/s if communication and computation are
exactly overlapped.\footnote{Our model is robust and numerically matches detailed profiling on nodes
  that actually perform at 10 TF/s for our working code (four Nvidia A100 four Mellanox HDR 200 Gbit/s cards), delivering 180 GB/s of bidirectional bandwidth to our
  application, and overlapping communication with computation almost exactly.}

With a large cache capacity one projects that for standard Krylov algorithms, the interconnect requirement is proportional to the local \emph{cache} bandwidth.
The coefficient is $\frac{1}{L}$ and typically $O(\frac{1}{16})$ which can lead to exceedingly high interconnect requirements, that are comparable to local high-speed memory bandwidths.

Algorithmic research in advanced MCMC sampling algorithms that both avoid critical slowing
down and improve the performance on realistic computer networks is required to deliver
many of the physics results that support the experimental program discussed above.


\subsection{Sample simulation parameters and costs}

In addition to the above common physics goals, 
we discuss some specific ambitions and computational costs for simulations with chiral (or domain wall) fermions.  The ambitions for improved staggered-fermion simulations are very similar.

Present chiral fermion lattice calculations use simulation volumes up to $96^3\times 192$ and use the most powerful supercomputers presently available to the Department of Energy.  Improved staggered-fermion simulation volumes are currently as large as $144^3 \times 288$ and are expected to grow to $192^3 \times 384$.

Our physics goals require calculations with ensembles of
gauge fields with physical volumes large enough to ensure that finite-volume effects are
under control. Such simulations require increased lattice sizes.

A number of specific simulations, Table \ref{tab:costs},
have been proposed with estimated costs
in a Snowmass white paper~\cite{NormanSnowmass},
and the same methodology can be used to estimate
the requirements of the ideal ensemble for flavor physics.

The final entry
is associated with
physics in the B-meson system indicated
in Snowmass white paper~\cite{OliverSnowmass}.
 A $256^3\times 512$ lattice at a lattice spacing $a = 0.04$ fm ($a^{-1} \sim 5$ GeV) would allow us to simulate up/down, strange, charm, and bottom quarks at their physical mass in
a 10 fm box with $m_\pi L = 7$.

These simulation goals clearly demonstrate a need
for computers at least 10x more capable than the coming Exaflop
computers. Since the performance is required to be delivered
on a real-code performance basis, and efficiency will not be 100\%,
more than an order of magnitude improvement, perhaps, from both
algorithms and computing are required.

\begin{table}[hbt]
\begin{center}
\begin{tabular}{|c|c|c|}
\hline
Lattice volume & $a^{-1}$ GeV & Exaflop hours\\
\hline
$32^3\times 64$	&1.4&	1.5\\
$40^3\times 96$	&1.7&	3.5\\
$48^3\times 64$	&2.1&	7.5\\
$48^3\times 96$	&1.8&	7.54\\
$64^3\times 128$&	2.4&	25\\
$96^3\times 192$&	2.8&	120\\
$64^3\times 256$&	2.4&	50\\
$96^3\times 384$&	2.8&	250\\
$128^3\times 512$&	&	1500\\
\hline
$128^3\times 512$& 5.0	&	12000\\
\hline
\end{tabular}
\end{center}
\label{tab:costs}
\caption{Proposed lattice volumes and cost estimates 
in \emph{sustained} Exaflop hours, scaled from current
simulations on Cori (NERSC) and Summit (ORNL).
Volumes and estimates are proposed in Snowmass white paper~\cite{NormanSnowmass}, while the final 
entry uses the same methodology to estimate the cost
of the most expensive proposed B-physics capable simulation.
}
\end{table}

\section{Computational landscape and challenges}
\label{Sec:computational-landscape}

We have demonstrated above that there is a 
deep need for future supercomputing that reaches far 
beyond the Exascale over the next decade.
Since Moore's law is hitting scaling limits, it is important
to consider whether prospects for greater computing power are realistic.

In this section we discuss the opportunities presented
by some new technologies that give realistic prospects
for improving system-level design.
In particular we will consider new memory and packaging technologies
that may enable better \emph{recruitment} of expensive logic
silicon for useful computation. Three-dimensional memory and advanced packaging
offer the opportunity to make connections to memory 
in numbers governed
by microscopic rather than macroscopic constraints. In this transition there is ``plenty of room at the bottom'' for sustained gains to be 
made in the more efficient use of transistors 
as a response to slowing of Moore's law for transistor density.

DOE HEP computation is likely to remain composed of both ``high-throughput computing'' (HTC or capability) and ``high-performance computing'' (HPC or capacity).
HTC exploits the trivial parallelism of independent processing
nodes and independent Unix processes, and achieves high aggregate throughput
by running many work items in parallel.
High-performance computing has an enormous investment in message-passing-based interconnects. Almost all HPC software is written to the Message
Passing Interface standard (MPI) and this appears a good long term 
investment 
because it supports the asynchronous decoupling of long latency operations, such as 
remote data access, from local processor execution. 

Within each Unix process, the elements of the  
computation performed by both HPC and HTC may or may not be
amenable to \emph{acceleration}.  There are a variety of acceleration
options, the most frequent of which are Graphics Processor Unit (GPU) accelerators (typically from vendors such as Nvidia, AMD or Intel). 

The principal technical difference between a HPC capable system and a HTC computer is the performance of the 
interconnect and potentially the degree of vendor specificity of the
software environment; for example whether the computing nodes are on a closed network among other issues.

While exotic hardware can seem attractive, 
it is worth remembering that at the time of writing, the fastest computer in the world is the RIKEN Fugaku system, which is based on a standard ARM multi-core
CPU architecture that is commonly used in mobile phones, combined with an integrated network
and high bandwidth memory. It has vector-computing acceleration,
but does not make large compromises in either software or special-purpose directions.

Some simple physics explains much of
recent and presumably future 
computer architecture trends. This enables us to attempt some degree
of informed 
forecasting, perhaps even over a ten-year
timescale, as we discuss in the following subsections.

\subsection{Wire-delay considerations}

Wire delay is the amount of time for a signal to propagate down a wire, and is proportional to the ``discharge time constant'' $RC$, where $R$ is the resistance and $C$ the capacitance.
We can model a wire as an isolated semi-infinite rod of metal of volume
$L \times \pi r^2$ and look at the scaling of resistance and capacitance as
all dimensions are ``shrunk'' by the same scale factor.
This gives rise to the discharge time constant,
$$RC \sim 2 \rho \epsilon \frac{L^2}{r^2} / \log(r_0/r) \sim  \mathrm{const} \times \frac{L^2}{r^2},$$
where $\rho$ is the metal resistivity, $\epsilon$, the dielectric permittivity and $r, r_0 \ll L$ are length scales small enough that the potential remains well
approximated by the long rod approximation.

This is profound, because, to a great degree (ignoring logarithmic terms and finite size effects),
wire delay depends \emph{only} on the geometry or aspect ratio of the wire. Shrinking
a process by the same scale factor in all dimensions does not reduce wire delay. 
Wire delay places an irreducible floor below which \emph{faster transistors make no difference}. 

It is interesting to note
that, historically, process advances announced in the press have included materials changes designed
solely to move this wire delay floor out of the way of transistor scaling:
``copper interconnect'' (180 nm) and ``low-k`` dielectric (100 nm) improved $\rho$ and $\epsilon$ respectively changing the RC time constant. 
The balance of a timed circuit 
becomes worse as transistors become faster without changes to the
materials that control wire delay.

The immediate corollary of the aspect ratio dominance of wire delay is the following:

\begin{center}
\fbox{\begin{minipage}{0.8 \textwidth}
\emph{Multiple processing units with broad wire long-haul buses is
the only possible high speed wiring
strategy for a planar 10 billion transistor  device}.
\end{minipage}
}

\end{center}

There must be a low number of long-range ``broad'' wires (bus/interconnect) since they consume area,
and a high number of short-range ``thin'' wires giving densely connected regions (cores).

One might note here that the mean wire length \emph{could} be reduced with devices containing
multiple logic layers of transistors, but this would require \emph{radical} changes to silicon
manufacturing and cooling technology to obtain truly three-dimensional high logic, whether by
chip-stacking or single-silicon-chip lithography. One might hope that, in future, current
thermal issues can be solved and that
silicon cubes can add a radical new dimension to digital computing.

\subsection{Hierarchical memory and multiple address spaces}

One important application of the previous simple analysis of wire delay arises when we consider the connection of a calculation device to memory chips. 
Long copper traces across a printed circuit board fall into precisely
the worst case corner for wire delay. Further, given the speed of light and GHz-scale clock rates,
distances of order 30 cm (= 1 light-nanosecond) set the length scale at which transmission-line behaviour sets in.
These considerations and, at high speed and with long wires, also, transmission line
dynamics, impose constraints on memory technologies.

Data motion is often the single greatest performance and power
bottleneck to be addressed in computation. A nuanced approach to analysing algorithms
will distinguish data references by their most likely origin (e.g. cache, local memory, remote node).
For a bandwidth analysis, the ratio of the floating point instructions executed to the bytes accessed
for each of the cache, memory and interconnect subsystems is called the \emph{arithmetic intensity} of the algorithm and is a key classification to understand if future computers are engineered to support efficient execution of the algorithm.

Since the cost of floating point units has decreased exponentially, while the cost of macroscopic
copper traces has remained (relatively) static, it is often the case that execution is memory limited. Many scientific algorithms can be categorised as either bandwidth or FPU
bound, for example, using the Berkeley roofline model~\cite{roofline}, and as discussed
above, the single-node performance of QCD codes are typically bandwidth bound with
the L2-cache bandwidth being dominant for multiple right-hand sides, 
while either the network bandwidth and/or memory is the rate-limiting step in gauge
ensemble generation.

\begin{figure}[hbt]
\begin{center}
\includegraphics[width=0.5\textwidth]{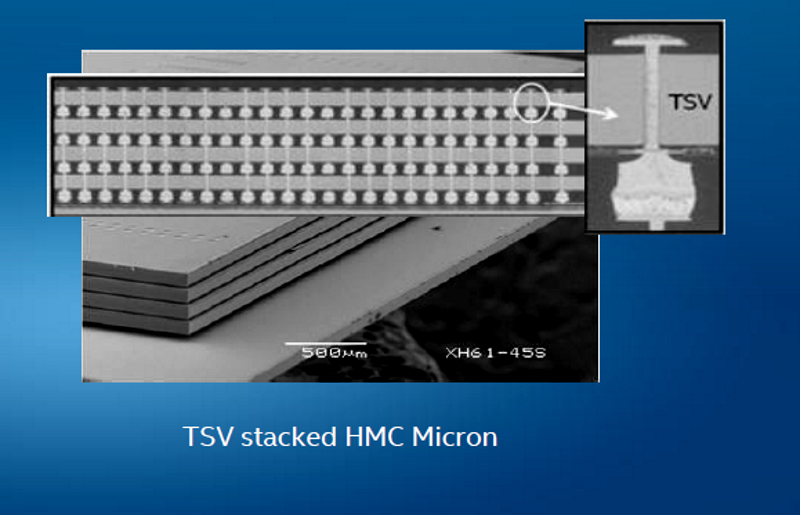}
\end{center}
\caption{ \label{fig:TSV} A micrograph~\cite{EETIMES} of a through-silicon via bus structure giving thousands of bit lanes
connecting memory chips with using a stacked configuration to give wiring geometries favorable from an energy and delay perspective.
}
\end{figure}

In order to substantially address memory limitations it is necessary to change both the aspect ratio of the wires involved, and vastly increase the number of wires that can carry data concurrently.
Fortunately the computing industry has already developed technology directions responding to this imperative with the 3d ``High Bandwidth Memory'' (HBM) technology. It is becoming a widespread solution.
These chips contain memory layers with large numbers of vertical metal rods (through silicon vias or TSV's) 
used as high-bit-count buses skewering the device, Figure~\ref{fig:TSV}. 
The ability to stack such chips vertically, with solder 
connections between TSV's, gives a favorable aspect ratio,
low wire delay and power. 
Furthermore, the miniaturization enables a high bit-lane count connection that 
gives a massive data rate for reading memory pages;
it is exciting that after a long period of conservatism in memory design, 
there is no clearly defined barrier to further increases in the number of bit lanes and bandwidth.

\begin{figure}[hbt]
\begin{center}
\includegraphics[width=0.5\textwidth]{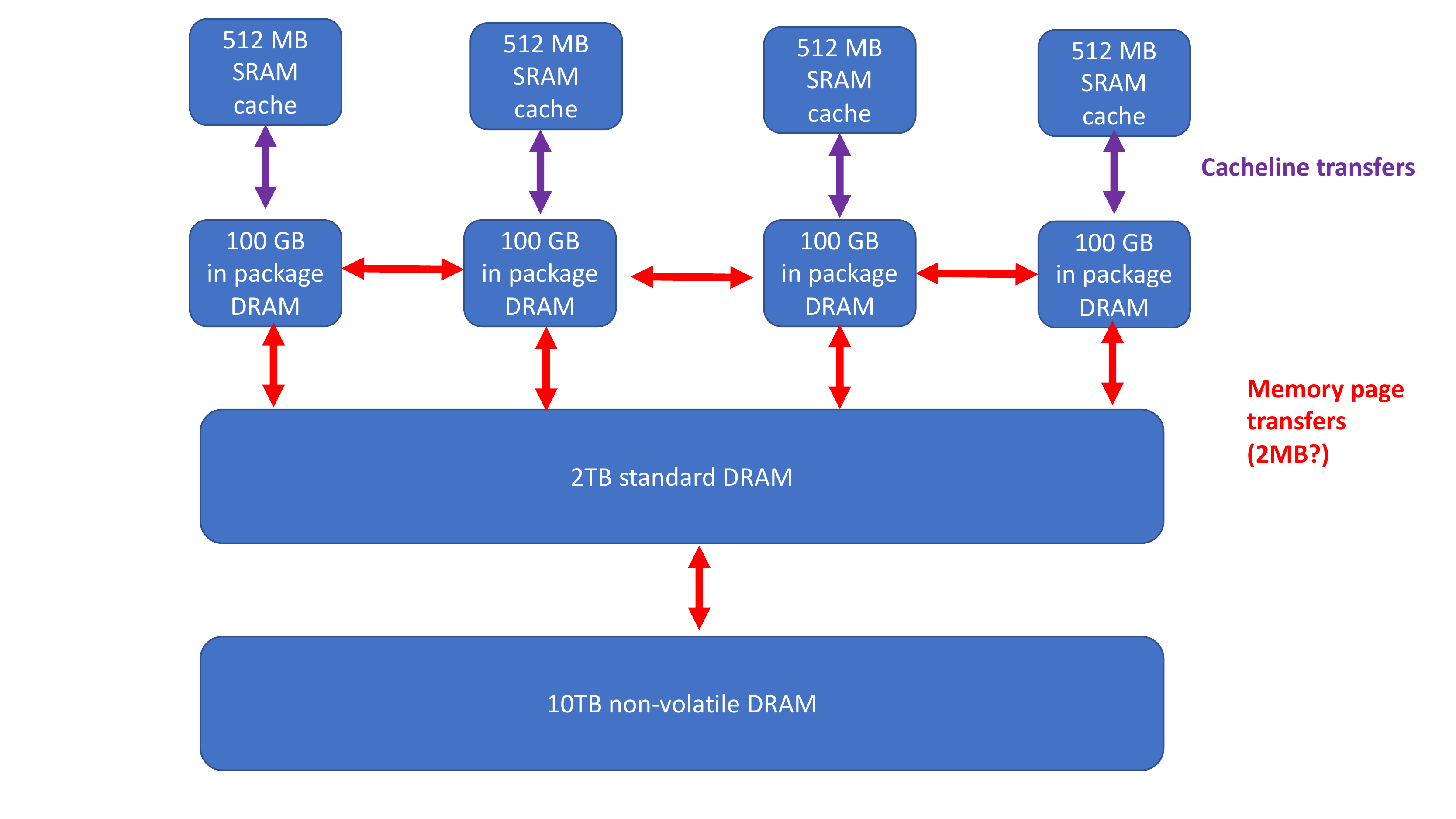}
\end{center}
\caption{ \label{fig:Memory} Independent of the underlying computational
technology (cores, GPUs, FPGAs etc), the underlying physics of
electronics is driving a hierarchical computing node memory system organization in the direction of accelerators.
}
\end{figure}

The advent of HBM is driving a rather revolutionary development across the industry towards on-package integration of memory, 
where a sizable volume of memory is physically co-located with the processing unit.
This may be done either through 2.5D integration with memory stacks placed along side a processing unit on an interposer (i.e. a miniature circuit board), or full 3d integration where memory chips are bonded on-top of a computational processing unit using TSV's. Recent Nvidia GPUs have used a silicon
interposer and recent GPU and EPYC CPU designs from AMD are multichip modules.
Intel has developed an ``EMIB'' multi-die package technology that enables bridging between adjacent die~\cite{EMIB}.
Rather than using a large silicon interposer,
this links together neighboring die in a multichip package much as
a child might join together two 8x8 LEGO squares to
make a 16x8 surface using a small 2x2 backing block.

A further significant emerging memory technology is a novel form of non-volatile phase-change memory, typically based on an amorphous/crystaline glass cell. This should enable increased memory density, and in particular is suggested to support many bit-cells
vertically stacked in multiple layers, promising four times higher density than DRAM. 
The most exciting application lies in plans
to give such devices conventional DRAM electronic interfaces.
It is a reasonable assumption that once a page is open, the column accesses
should be no slower than in a conventional DRAM since the electronics will be similar, while page-miss latency may be larger. 
Certainly, there are
already standards for non-volatile memory modules, and
this is likely
a disruptive technology for large memory algorithms, such as the use of Dirac eigenvectors and the assembly of multi-hadron correlation functions.

It is reasonable to assume that on a decade time frame, memories will  remain hierarchical and likely become even more hierarchical than at present. 
Based on the above arguments there are good fundamental electronic reasons to have computational devices with the following properties:
\begin{itemize}
\item one hundred or more independent processing units.
\item internal caching with several to many times local memory bandwidth.
\item local in-package memory with several TB/s high bandwidth memory ranging to 100 GB scale or more. 
\item slower conventional DRAM-based system memory with greater capacity (multi TB) and a few 100's of GB/s bandwidth.
\item Non-volatile memory with even greater capacity, longer latency and potentially reduced bandwidth.
\end{itemize}

Such a memory organization is illustrated in Figure~\ref{fig:Memory}.
In this diagram the red data transfers could either be managed by user
software, or by demand-driven fetching (at the cache line or memory page level
of granularity). This organization is potentially independent of the
computational technology. For example, low power multi-core CPUs with high
bandwidth memory could be used as accelerators for multi-core CPUs with
large slow memory.

Given such hierarchical memory, there is a significant question about how this should be
managed. There are two logical solutions, and several means of implementing each.

The first is the ``easiest'' from a computer vendor perspective: to solve the
problem with application and library software. 
One can expose these as
multiple address spaces and have the programmer or programming interface handle data placement.
This is a familiar element of the (now) traditional GPU or ``offload'' model, which
can be implemented by giving the programmer two types of pointer: 'host' and local
'device' pointers, and requiring the programmer to keep track carefully of what data is stored where.

Another common software solution, taken by SYCL~\cite{SYCL},
is to break arrays into distinct
data-storage containers and accessors, with the
process of gaining access implicitly triggering data motion.

The second and more elegant solution is to solve the problem with \emph{operating system software}.
This presents a single logical address space in Unix \emph{virtual memory}, but moves physical
pages of data to a nearby location as and when required by a computational unit.
This is a generalization of the familiar Unix virtual-memory swapping mechanism and
could in future be systematised to cover even more hierarchical memory systems including
both volatile, non-volatile, HBM etc... 
Such ``unified'' address space solutions presently introduce some performance overhead, but
greater flexibility of memory page sizes to amortize transfer overheads would surely make this
a good, long-term solution that simplifies the programming challenge. 

Unified Virtual Addressing, as used in CUDA, SYCL and HIP, is a good way to 
organize this and minimise the
programmer burden. This uses existing low latency
``translation look-aside buffer'' (TLB) virtual memory mechanisms
to search for a local copy of data. The default page in 
user space for HPC systems could
be promoted from 4KB to 2MB (or an intermediate page size introduced) so that 
paging overheads are amortised on large arrays. This viewpoint would
naturally extend to hierarchical systems comprising both slower non-volatile memory and faster conventional DRAM.

In the Fugaku computer, high bandwidth memory has been combined with
multi-core ARM processors, using HBM as the main memory.
In some cases, such as the Intel Knight's Landing, 
HBM has been used in conventional CPUs as a cache. However
with multi-gigabyte caches the status information (tags)
are also huge and must be stored in the HBM. This makes the 
process of checking for the data in the cache expensive. In contrast, a
virtual memory translation typically hits in the translation cache. Larger user space page
sizes would enable good coverage in translation caches
and good efficiency for page migration.

Page sizes as small as 4KB are almost certain
to lead to low performance in user space in future, and delivered
bandwidth will grow with the granularity of transfer. One might hope that
an improved virtual memory system would have an option to only use larger memory pages, such as 2MB in future to amortise kernel overheads.

The development of such a coherent virtual memory direction for hierarchical memory 
may require incentives from the Department of Energy ASCR 
leadership and could perhaps 
lead to significant benefits to computing environments.

\subsection{Interconnect technology}

\begin{figure}
    \centering
    \includegraphics[width=\textwidth]{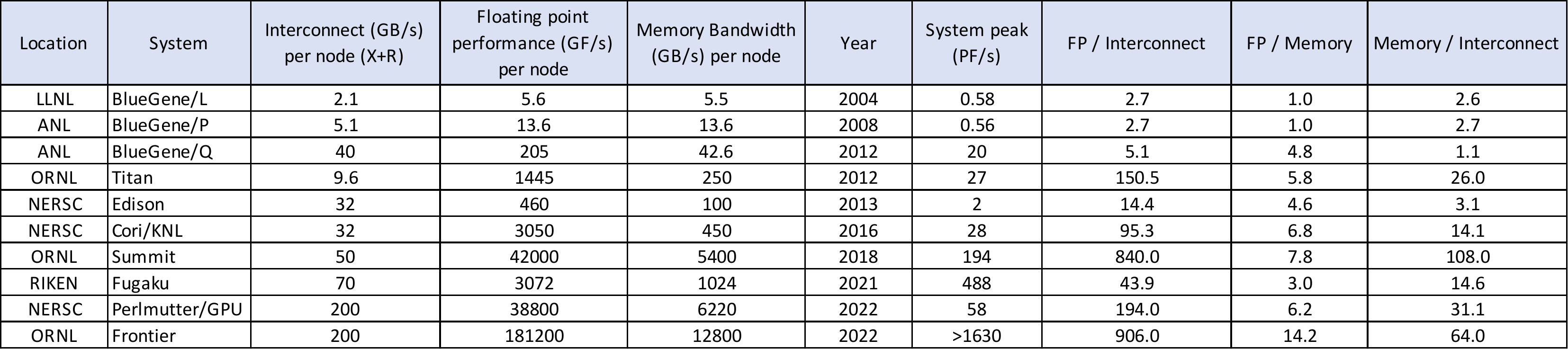}
    \caption{Evolution of high end HPC computer systems over the last two decades.
    The interconnect has only increased slightly in performance while larger increases have been
    seen in local memory bandwidth and particularly in (peak) per node performance with the 
    introduction of accelerated computing as powerful (and expensive) nodes.
    These are often accompanied by powerful memory networks interior to the node.
    This growing ratio of both local
    floating point and local memory bandwidth to interconnect performance leaves interconnect a bottleneck in many current algorithms for Lattice QCD ensemble
    generation.
    }
    \label{fig:network}
\end{figure}

From a technological perspective, the interconnect is the most significant lagging capability in computer evolution, as illustrated in Figure~\ref{fig:network}.
The ratio of memory bandwidth to local floating-point performance and the 
interconnect bandwidth to local floating-point performance has changed in the last twenty
years from numbers of order one, to numbers of order ten (memory) and order one thousand (network).
While it is true that these fat nodes are capable of larger problems, permitting a lower surface to volume ratio and, thus, a proportionally lower
network demand, it is still the case that parallel algorithms must be designed to avoid the substantial
cost of accessing remote data.  
We must invest in developing efficient algorithms that recognise that new computers have a series of weakly linked islands of high performance to realise the potential of computing ten years from now.
Fortunately the separation of length scales is a fundamental element of physics and effective
theories, and multi-scale algorithms are a reasonable goal for research investment.
    
We may also ask: is limited network bandwidth a technological necessity?
While 100 Gbit/s of copper trace in a larger printed
circuit board (PCB) will cost as little as pennies, the same
100 Gbit/s of bandwidth in copper cables costs just under \$100 USD. A similarly performing 100 Gbit/s active optical cable over
four bit lanes costs of order \$1000 USD.

The arguments given in the above discussion on 
wire delay and transmission line behavior
are \emph{not} applicable to fiber-optic cables. Modern interconnect cables that exceed
roughly 2 m in length are predominantly fiber optic. 
They increase costs, as optical 
transceiver technology uses exotic non-silicon 
processes (that differ from standard logic silicon), and cables (such
as QSFP56) have electrical interfaces with active optical transceivers on either end. This raises price and power substantially.

The stated purpose of very large-scale integration (VLSI) is the 
reduction of cost by combining
multiple ``parts'' in a single chip. 
Significant industrial effort is being focused on 
technologies such as silicon photonics to 
implement fiber optic transceiver logic in 
a standard silicon process. Combined with vertical-cavity surface-emitting laser (VCSEL)
technology there is a real possibility of a chip-to-chip light path, perhaps multiplexing
laser cells with different colors, with greatly increased
bandwidth removing active optical transceivers and alleviating bandwidth limitations.
However the appetite and focus of industry on this direction may depend in large part
upon Department of Energy requirements for future supercomputers.

In notable cases, such as the IBM BlueGene computing line, system on a chip technology
has enabled integrating not just network transceivers but \emph{also} a distributed
router element. This enables the integration of not only a network interface but \emph{also}
the switch system in the node silicon, substantially reducing the 
added cost of the high performance
interconnect from over a thousand dollars per node to dollars per node. 
At the time of introduction
BlueGene/Q had bandwidth equivalent to eight contemporary 
infiniband cards but connected nodes together
within a rack with glueless wiring (i.e. no intermediate chips). 
Optical transceivers
were used on the surface of a computer rack containing 1024 nodes, suppressing the cost
of optics in the system. This example emphasizes the importance of Department
of Energy ASCR leadership in encouraging innovative system integration.

\subsection{Processing Units}

At the core of a computing system is the processing unit, defining the
interface a programmer controls directly, if at least through a compiler, and which accesses the memory to perform its assigned tasks.

The traditional programming unit is the microprocessor, which is commonly called a
processor ``core''. They present a logical state and instruction fetching and
execution. They are mature and typically highly optimised for the highest
performance for (cleverly giving the illusion of) serially executing a 
single sequence of instructions. Since processor cores optimised
in this way are large and power hungry, it has become common to improve floating
point throughput and power by using short vector ``single instruction multiple data'' (SIMD) units to increase the density of floating point hardware.
However, this design does introduce constraints and not all HEP computing can
benefit from such hardware. Lattice QCD \emph{does}, and it
can make use of all current SIMD architectures and even longer vectors would
be useful.

Modern GPUs provide powerful alternatives to CPUs and deliver excellent performance and power performance for
a number of reasons. Firstly, an accelerator architecture may be more specialized than processor cores that target the best general purpose
single-thread performance (and a full range of features). Secondly, by using a private memory system,
more aggressive technology decisions may be made. A host CPU is retained for executing general code that is not well handled by
a GPU thread, and an \emph{offload} model is used where critical loops and data are marked by software for execution and placement
on the accelerator device.

All GPUs at this time present a parallel multi-dimensional (1d to 3d) loop as the primitive looping construct.
Similar to the Connection Machine computer, the machine model is that each instance of the parallel loop
body is presented as a different virtual machine or thread. However, syntactically, the implementation is less
elegant without data-parallel expressions in the high level language. 

Although GPUs are fundamentally SIMD architectures, addressing modes and masked execution are cleverly used to obscure this
fact and present a scalar processing model to the programmer, called ``Single Instruction Multiple Thread'' (SIMT).
In SIMT, a single instruction fetch unit controls \emph{multiple} logical threads, typically a number of O(32).
When some threads choose yes, and other threads choose no the divergence leads to loss of parallel throughput.

Accesses to thread-private data (stack, local memory and what the programmer would think of as local variables - if not in registers)
are addressed in a way that efficiently interleaves accesses to corresponding local memory locations by each thread in
a physical memory array.
One might imagine that electronically the ``thread'' index within a parallel execution group dictates the byte address within
a hardware SRAM data bus. This ensures that when a group of software ``threads'' concurrently execute the same instruction, they
all access the matching variable on their respective stack or local memory. The accesses will be transferred as a physically contiguous
data transaction even though the virtual addresses are relative to different stack pointers or in a local memory space.

Both GPUs and CPUs make use of forms of ``hyperthreading'' but with varying degrees
of expected parallelism. CPUs are very much focused on single thread performance
(i.e. time to solution for for single serial calculation) while GPUs are rather more 
focused on parallel performance or throughput. Modern CPU processor cores will concurrently
execute instructions from between one and four independent processes or threads, each
of which maintains its own instruction pointer and program state.
In contrast up to 32 independent instruction pointers are common in current GPU processors. 
The low thread count of the CPU makes for
smaller register files with lower latency to access, while GPUs use larger and multi-banked
register files with potentially longer latency. In this way,
a GPU is able present a greater
number of independent outstanding loads to the memory system with greater throughput
from the additional \emph{memory level parallelism}. This parallelism is easier to 
find and manage in the GPU since the independent threads are trivially parallel while in the CPU direction significant engineering resource
is devoted to dependency tracking.

One might imagine that conventional microprocessors could, in principle, add addressing modes that facilitate a similar SIMT model in their
vector extensions and become more GPU-like to make vectorization easier.
It is also possible low power CPU multi-core designs coupled with GPU
style high bandwidth memory systems could provide viable alternatives 
for some of the HPC workload, particularly those which do not map well
to the data parallel model. The very large peak performance of GPUs is
relevant to dense matrix operations, but bandwidth limits the performance
of many other common algorithms. The success of the Fugaku computer combining
low power Fujitsu/ARM processor cores with 3D memory and an integrated network is a good example.

\subsection{Machine Learning and non von Neumann hardware acceleration}

Fixed-function acceleration
is possible in the area of machine learning (ML), or more generally
problems involving dense matrix multiplication,
with tensor cores in GPUs and other hardware specific
to large matrix multiplication. These are commonly on 
64-bit and 32-bit floating point arithmetic, or even 
on 16-bit floating point arithmetic. Mixed precision functional units are common. 
Fixed function dense matrix
hardware such as tensor cores have proven difficult to use because the 3x3 complex 
gauge link matrices do not map naturally to such hardware.

Reduced precision formats using non-IEEE mantissa and significand
have been found to be useful
in Machine Learning including non-IEEE 16-bit~\cite{intel-patent,bfloat} and 8-bit arithmetic~\cite{Hopper}. A joint patent for a 
hardware implementation for the modified
16 bit floating point format
was submitted by one of the authors (PB)
with Intel in 2018~\cite{intel-patent} wherein it was
demonstrated to provide better support for training in machine-learning.
Lattice QCD can also make use of reduced-precision arithmetic, and gains
64-bit precision accuracy using 32-bit or even 16-bit precision floating point
arithmetic in a preconditioner or mixed-precision solver\cite{QUDA}. 

Machine learning can broadly be decomposed into ``training'' and ``inference'' phases. Training often involves scientist supervision
for parameter tuning with time to solution being critical. 
Distributed machine-learning training parallelizes the training process.
The use of small ``mini-batches''
in stochastic gradient descent leaves the problem very sensitive to 
Ahmdahl's law slowdown. In the massively distributed
limit~\cite{Kurth}, the machine-wide reduction of a gradient vector
calculated across many randomly selected training samples in
a minibatch is a significant performance bottleneck~\cite{Baidu}. 
Here the extreme interconnect requirements of QCD and 
distributed machine learning align~\cite{Boyle:2017xcy}.

Inference with a previously trained network is typically
performed on many independent events of data, and is in many cases a
trivially parallel or HTC problem. However, when inference is used
in either hard or soft real-time situations (as occurs in HEP experiment)
the deadlines impose challenging performance issues.

Accelerators from GraphCore and Cerebras are good examples of
ML-optimised accelerators in
addition to more standard GPUs by Nvidia, AMD and Intel. 
Reconfigurable computing hardware 
such as Field Programmable Gate
Arrays (FPGAs) or spatial accelerators are powerful
options for certain algorithms. 
They may be accessed via a dataflow
model where a hardware circuit 
is configured in a non von Neumann approach
to stream data from memory, calculate and store results in memory.
They may be particular effective in eliminating software latency 
in real time inference environments.

\subsection{Architectural diversity}

The overview above shows that the computing landscape at this time displays a
wonderful proliferation of computer architectures.
It opens vibrant and healthy opportunities for innovation and competition and
for the introduction of new and competitive ideas.
However, it also risks introducing special-purpose or
less-general features than with previous systems. The memory spaces
are often fragmented with different types of pointer and location of
data. There are multiple instruction-set architectures in use, each
with different programming interfaces or models.
A single program frequently has different segments that target different
types of instruction sets within a single computer node. Indeed, 
some of the computing models available (such as FPGAs or spatial
architectures) are not von Neumann computers. Others have dedicated
fixed-function acceleration of matrix-matrix multiplication for machine
learning.

This diversity presents a significant challenge to the scientific programmer
with either legacy code or the need to use more than one system.

\section{Software challenges}
\label{Sec:software-engineering}

Text-book computer engineering~\cite{HennessyPatterson} suggests that code optimizations should
expose an admixture of two forms of locality: spatial-data-reference locality, and temporal-data-reference locality.
Buses are used that transfer many contiguous words of data at a time in many levels of a computer memory system.
For example, when a DRAM page is opened (row access), very many consecutive bit cells are read and \emph{page hits} (column access) incur much less
overhead than would occur with a new random access, which would give rise to different latencies. Further, multiple layers of caches are used; the access bandwidth and latency
characteristics become poorer the further one travels from the processor core while the storage capacity correspondingly increases.
The transfers between levels in the memory system are performed in aligned chunks called cache lines, with sizes $\in\{32,64,128,256,512\}$ bytes
depending on the level and computer architecture. For example, modern Intel chips typically use a 64~B cache line in all layers of the cache hierarchy, while
the IBM BlueGene/Q system made use of a 64~B L1 line size and a 128~B L2 line size.

Spatial reference locality arises because memory systems are fundamentally \emph{granular}. If one accesses one word
of a cache line, the entire cache line is transferred over the (rate limiting) buses in the system. In order to maximise performance, algorithms and codes should ensure that they
make use of all the data in cache lines that are transferred, by laying out data to give spatial locality of reference.
Architectures use cache lines based on the observation that many algorithms (such as linear algebra) have such access patterns.
Temporal locality of reference arises when data referenced is likely to be referenced again soon. Thus caches store recently accessed
data.
Both of these hardware optimizations are relatively easy to exploit by writing software that loops or block in the right order.

Additional considerations apply when working with architectures that implement short vector instructions (single instruction, multiple data or SIMD). Particularly when the processor core is complex, these instructions represent one of the cheapest ways to to enhance peak floating point performance, as it vastly cheaper to build hardware to execute a single arithmetic instruction or load on a contiguous block of N elements of data than to generally schedule N independent and decoupled instructions. In order to exploit these instructions, code optimisations should expose what one might call spatial operation locality. Although there are obvious applications in array and matrix processing, even matrix transposition shows that it is surprisingly hard to exploit SIMD in general because you must arrange to have same operation applied to consecutive elements of data.

\subsection{Programming models}

There are a number of different programming models
one must consider in the present computing landscape. The following can all in principle
be combined with MPI message passing.\\ \\
Present LQCD ECP codes support:
\begin{itemize}
    \item traditional CPU core, SIMD floating point instructions, and OpenMP threading,
    \item CUDA-based GPU programming,
    \item HIP-based GPU programming,
    \item SYCL-based GPU and FPGA programming.
\end{itemize}
Expected to be supported by future LQCD software:
\begin{itemize}
    \item OpenMP 5.0 offload GPU programming,
    \item C++ parallel STL offload programming.
\end{itemize}
The challenge of writing high-performance and portable code is three fold:
\begin{itemize}
\item syntactical differences, 
\item semantic differences, and
\item data placement.
\end{itemize}

Firstly, 
the syntax for offloading loops depends on the underlying software environment. 
Of the Department of Energy open-science pre-exascale and exascale systems 
(Summit, Perlmutter, Frontier and Aurora), all are accelerated
with various architectures and there are three
distinct native vendor programming models: Nvidia/CUDA, AMD/HIP and
Intel/SYCL.
One approach may be to identify a suite of abstractions that are both compact and 
adequate to write portable and high performance software with 
a single abstract interface that can itself target all of the above.
Techniques introduced by the RAJA~\cite{Raja}
and Kokkos~\cite{Kokkos} libraries use C++ device-lambda function objects 
to capture sequences of offloaded code.
Within Lattice QCD, a similar effort has been undertaken in 
the ECP project on the Grid software library~\cite{Boyle:2016lbp,Boyle:2017gzg,Grid2021}
and on the QUDA software library~\cite{QUDA}, which have both been ported to execute
under CUDA, HIP and SYCL.

Alternately, one could rely on a single standard interface that is used
to provide portability. OpenMP 5.0 offload is potentially such an interface,
and may in fact be the appropriate portability layer in the long term.
OpenMP has already been transformative in making multithreaded programming accessible to the average scientific programmer.
However, there are other standards-based alternatives: SYCL is \emph{in principle}
cross-platform portable, but whether the performance of either SYCL or OpenMP reproduces vendor-native compilers giving \emph{performance portability} remains to be seen. Similarly, C++ parallel STL offers a more standard C++ interface to
offload and parallelism, but this is even newer than either SYCL or OpenMP offload.
Parallel STL presently has support from only one GPU vendor and, although a standard, the roadmap
to de facto portability remains unclear. 

At this time, abstraction and support for multiple
programming models is likely the safest option for portable efficiency on near
term computers. For projects that wish to reuse interfaces to acceleration,
both Kokkos and RAJA provide good options as Department of Energy ASCR projects.

Secondly, the somewhat larger challenge is to write a single program that captures the differing semantics between SIMD and SIMT execution models. 
The optimal data layout changes with the parallelism model. Both SIMD and SIMT use underlying vector architectures
  and a partial ``struct-of-array'' transformation is needed in data arrays in memory.
  However they semantically differ in the behavior of local variables within functions.
  In a GPU, each ``lane'' of the underlying SIMD executes a different logical instance of the same function, and thus processes
  scalar items, while in a CPU local variables remain (short) vector data types.
  Optimal software cannot be invariant when the architecture is changed, and rather
  to target both efficiently it is necessary to design a programming style that transforms covariantly with the architecture.
  Some QCD implementations use compiler intrinsics to obtain performance, 
  however software could be substantially simplified
  if (perhaps working with the Kokkos team) we could successfully advocate
  that C++ ``std::simd'' support
  vectors of complex number datatypes and standardise permutation and vector interleaving patterns.

Thirdly, placing data and managing data motion should be simple and even transparent. 
Marking up every loop with 'copy-in' and 'copy-out' pragmas is laborious, for example. As advocated above, unified virtual memory and a systematic approach to
increasingly hierarchical memory by the computing industry would give a significant and long term
benefit to scientific programming. 

The Lattice QCD recipe for building portable code on accelerated systems could be summarised as:
\begin{itemize}
\item abstract offload primitives and device function, attributes,
\item abstract memory allocation primitives,
\item abstract communication patterns such as halo exchange, shifting, reductions and sub-block reductions.
\end{itemize}
Unless relying on unified virtual memory or similar:
\begin{itemize}
\item Software managed device cache for host memory regions,
\item Distinguish accessors (views) of lattice objects from the storage container.
\end{itemize}
To obtain portable performance:
\begin{itemize}
\item Abstraction capturing SIMT and SIMD models in a single interface,
\item Provide Stencil differential-operator-assistance primitives: neighbour indexing, halo-exchange patterns. 
  \end{itemize}

This has been achieved in at least two major code bases for the 
HIP, SYCL and CUDA APIs. One of these also gives good performance
on several CPU SIMD architectures.
Newer interfaces like OpenMP 5.0 offload and ISO Standard C++
(including parallel implementations of the C++ Standard Template Library, also known as Parallel STL or pSTL )
will be adopted as and when enabled
both by the maturity of the respective standards and of compiler implementations.
In particular, compilers are only now starting to adopt OpenMP-5.0 features and
ISO C++ parallelism is implemented in several compilers for CPU architectures and even by one GPU vendor.
However, the C++ standard does not currently feature the concept of different memory address
spaces and hence GPU implementations rely on automatic data movement and are
currently limited to parallelism on one type of device (either the host CPU or GPU)
via the STL at a time, requiring other frameworks in a hybrid application.

One hopes that this might consolidate the
proliferation of programming interfaces as the lack of standardization
imposes a \emph{very significant software development burden} on the science community with
duplication of effort for multiple systems. This software development
underpins the entire community effort and requires support to make the Snowmass science goals feasible.

\section{Ecosystem Challenges}
\label{Sec:ecosystem-challenges}

For this ambitious physics program to be viable, a
significant computational challenge exists, in order to enable a series of calculations on ensembles from $a^{-1}=3-5$ GeV, and lattice volumes
from $96^3\times 192$ to $256^3\times 512$. The challenges exist on multiple
fronts: intellectual in developing algorithms that evade critical slowing down, software engineering to develop well-performing and portable code on an evolving range of supercomputers and programming models, and technical to remain engaged with the DOE HPC community as systems are planned and developed.

USQCD has had senior-staff, post-doctoral researchers, and post-graduate students funded under the Exascale Computing Project (ECP) and the SciDAC program.
 The ECP project includes algorithmic programs in ``Critical Slowing Down'', and ``multilevel solvers''~\cite{Richtmann:2019eyj,Yamaguchi:2016kop,Joo:2019byq}. It has also funded the development of high performance software portable to Exascale hardware~\cite{Grid2021,Boyle:2016lbp,Boyle:2017gzg,QUDA}.
 The community activity is dependent on the existence of
bespoke software environments that enable
efficient simulation on rapidly evolving supercomputers,
which requires significant expertise to develop and sustain.
 It is important that these gains continue to be realised, with continued funding throughout the Snowmass period, or an opportunity comparable to the available gains in computer performance will be lost. 
 
 It is important that flexible high-performance software is developed for a diverse range of architectures that tracks the DOE computing program.  The life cycle of scientific code is at least 10 years, and 
the health of a community depends on large code bases (up to 200,000 lines of code) which do not have a 
secure model for development and support which places investments at risk. 

A worrying trend is also identified in a recent preprint \cite{2022arXiv220302544R},
which notes that innovation in the design of computing devices is now much more heavily
influenced by the needs of mobile-phone companies and cloud data centers and their associated services
than by Governments due to the enormous financial power wielded by the leading corporations
compared to levels of investment which Governments can bring to bear. The paper also
notes ``Talent is following the money and the opportunities, which are increasingly in a
small number of very large companies or creative startups''.
Such a trend may continue to exacerbate retention challenges in the future.

\subsection{Lab supported software development}
Just as the large experiments require talented permanent staff
at the Labs to engineer experiments and manage sophisticated
long-term software systems, cost- and people-effective lattice
QCD desperately requires that career paths be created to retain
some of the most talented experts in software and algorithms.

The permanent staff will address HPC architectures as they emerge under the Computational Frontier. We aim to develop performance-portable high-level data-parallel code, with a write once and run anywhere approach that many domain scientists can modify effectively.
This effort must track the evolution of computing.
A key element of managing the science program is the early engagement with DOE HPC laboratory sites during the development, and years prior to installation, of major new facilities.  The lead time for porting to new architectures lies in the region of multiple years, and early engagement with emerging architectures
is required to ensure timely scientific exploitation.

\subsection{Joint Lab-University Tenure Track Appointments}

We believe the DOE should 
seek to foster the continued development of intellectual leaders
in computational quantum field theory.  This is now done very
well at the Labs, with a large fraction of the leading personnel
located there.   The health of the field requires a similar
cohort of individuals at the best universities, reflecting the
intellectual vigor and potential of this area to
contribute to DOE scientific goals. The creation of
such positions can be stimulated by DOE-funded joint, five year, tenure-track appointments. 

Theoretical particle physics is one of
the last area of physics to recognize the importance of
computation in forefront research and continued effort is
urgently required to overcome this historical bias, and
create a vibrant pool of skilled young faculty, and around 
them their PhD students and research groups.

\section{Conclusions}

We have discussed the importance of Lattice QCD to key elements of the future experimental
physics program. Theoretical results for 
hadronic parameters are critical inputs in 
searches for new physics: for example, the DUNE and Fermilab $g-2$ experiments, in addition to continuing searches in flavor physics.

Lattice QCD is uniquely dependent on high-performance computing.
High performance computing will likely continue to evolve rapidly, 
bringing both a massive potential for science, but also
posing significant challenges to exploitation and delivery of this potential.
This evolution must be coupled with continued innovation in
developing multi-scale-capable algorithms to solve the
challenge of critical slowing down in Markov-Chain Monte-Carlo and 
in multi-scale linear solvers for all fermion discretizations.

DOE support to develop the software and algorithmic environment in which computational
quantum field theory can be performed is important to enable the rich physics plans of the community.
The DOE has an opportunity to foster the development of a cohort of 
talented LQCD scientists doing this work.

Due to the nature of leadership-class computing facilities, continued
access to institutional-cluster computing for USQCD is a substantial
benefit to productivity, allowing the long tail of many small projects
with modest computer requirements to be carried out using easy-to-program architectures with a low barrier to access.

Scientific productivity would be enhanced if the DOE used its 
influence on world supercomputing to standardize programming interfaces and ensure simple memory models that hide underlying hardware complexity.

\section{Acknowledgements}

PB, TI, and CJ have been supported by the U.S. Department of Energy, Office of Science, Office of Nuclear Physics under the Contract No. DE-SC-0012704 (BNL). F.J is supported by UKRI Future Leader Fellowship MR/T019956/1.
A.P. also received funding from the European Research Council (ERC) under the European Union’s Horizon 2020 research and innovation programme under grant agreements No 757646 \& 813942.
This research used resources of the Oak Ridge Leadership Computing Facility at the Oak Ridge National Laboratory, which is supported by the Office of Science of the U.S. Department of Energy under Contract No. DE-AC05-00OR22725

\bibliography{main}

\end{document}